\documentstyle[twoside,fleqn,espcrc2,epsfig,cite]{article}

\newcommand{\beq}{\begin{equation}}
\newcommand{\eeq}{\end{equation}  }

\newcommand{\bec}{\begin{center}}
\newcommand{\eec}{\end{center}}

\def\d{\delta}

\def\E{\mbox{e}^+\mbox{e}^-}

\newcommand {\NP}     {Nucl. Phys.\ }
\newcommand {\PL}     {Phys. Lett.\ }

\newcommand {\PRep}   {Phys. Rep.\ }

\newcommand {\ZfP}    {Z. Phys.\ }
\def\coll{{Collab.}}
\def\etal{{\it et~al.}}

\title{Local properties of local multiplicity distributions 
in hadronic Z decay
\thanks{Presented at the XXVII Symposium on Multiparticle Dynamics, 
September 8-12, 1997 Frascati-Rome, Italy.}}

\author{S.V.Chekanov\thanks{On leave from
Institute of Physics,  AS of Belarus,
Skaryna av.70, Minsk 220072, Belarus.} with W.Kittel and W.J.Metzger \\
\vspace{2mm}
        High Energy Physics Institute Nijmegen (HEFIN),
        University of Nijmegen/NIKHEF, \\
        P.O. Box 9010, 6500 GL Nijmegen, The Netherlands \\
\vspace{2mm}
for the L3 Collaboration
}

\begin{document}

\begin{abstract}
Preliminary results on local multiplicity fluctuations
in hadronic Z decays are presented.  The data were obtained 
using the L3  detector at LEP.
It is  investigated to what extent Monte-Carlo models, which are 
tuned to reproduce global event-shape variables  and single-particle
inclusive distributions, can describe the local fluctuations
measured by means of bunching parameters. 
\end{abstract}

\maketitle

%
\section{INTRODUCTION}
\label{int}
 
The investigation of  the evolution of the probabilities  
$P_n(\d )$ of detecting $n$ particles in ever smaller  
sizes $\d$ of  phase-space windows (bins) 
provides  detailed  information on  QCD multihadron 
production  beyond single-particle densities, 
without trivial constraints from 
charge-  and energy-momentum conservation. 
A deviation of  this
distribution from that expected for purely independent particle
production can be attributed to 
dynamical local multiplicity  fluctuations. 

The important quest behind such a study  is 
the understanding of the origin of short-range correlations between 
final-state  particles.
As a consequence of these correlations, 
the normalized factorial moments (NFMs) $F_q(\d)$
of the local multiplicity distribution $P_n(\d )$ exhibit 
a power-like increase with decreasing $\d$, namely  
$F_q(\d )\propto\d ^{-\phi_q}$ \cite{bp}, 
where $\phi_q$ are constants called intermittency indices. 
This phenomenon  reflects the peculiarity of 
$P_n(\d )$ to become broader with
decreasing $\d$. 
Since NFMs satisfy  the scaling property  
$F_q(\lambda\d )=\lambda^{-\phi_q}F_q(\d )$,
this is widely regarded as  evidence that 
the correlations exhibit  a self-similar 
underlying dynamics.

Local fluctuations in  $\E$-processes 
have already been studied is several experiments \cite{cllr}. 
The data do exhibit approximate  
power-like rise of the NFMs with a saturation at small $\d$.
The conclusion has been reached that such a phenomenon is 
a consequence of the multi-jet structure of events,
i.e., groups of particles 
with similar angles resulting
in spikes of particles as  seen in  selected phase-space projections.
Parton showers, fragmentation, resonance decays
and Bose-Einstein interference can all contribute to these
correlations.
It has been found that for the statistics used at that time
current Monte-Carlo  models can, in general, describe the data,
even without  additional tuning.

\begin{table*}
\setlength{\tabcolsep}{1.5pc}
\newlength{\digitwidth} \settowidth{\digitwidth}{\rm 0}
\catcode`?=\active \def?{\kern\digitwidth}
\caption{NFMs and BPs for the   distributions quoted.}
\label{tab1}
\vspace{0.1cm}
\begin{tabular}{lrrr} \hline
Distribution &  $P_n$ & NFMs & BPs \\ \hline
 &  &  &   \\
Positive-binomial & $C_n^Np^n(1-p)^{N-n}$ &
$\prod_{i=1}^{q}(1-\frac{i}{N})$ &
$\frac{q-1-N}{q-2-N} $ \\ [0.2cm]
Poisson & $p^n\exp(-p)/n!$ & $1$  & $1$  \\ [0.2cm]
Negative-binomial & $\frac{\Gamma(n+k)}{\Gamma(n+1)\Gamma(k)}
p^n(p+1)^{-(n+k)}$ & $\prod_{i=1}^{q}(1+\frac{i}{k})$ &
$\frac{q-1+k}{q-2+k}$ \\ [0.2cm]
Geometric & $p^n(p+1)^{-n-1}$ & $\prod_{i=1}^{q}(1+i)$ &
$\frac{q}{q-1}$ \\ [0.2cm]
\hline
\end{tabular}
\end{table*}

Recently, it has been realized that the factorial-moment
method poorly reflects the information content of local fluctuations,
since the NFM of order $q$ contains a trivial contamination from
lower-order correlation functions 
(see reviews \cite{rev}). As a result,  
rather different event samples can exhibit 
a very similar  behaviour of the NFMs.
The fact that subtle details in the behaviour of $P_n(\d )$ are missing,
together with the small statistics used, may be the reason    
why different Monte-Carlo models can reasonably  describe
the local fluctuations measured in $\E$ annihilation so far. 

Another shortcoming of the factorial-moment  
measurement is that in moving to ever smaller phase-space bins,
the statistical bias due to a finite event 
sample ($N_{\mathrm{ev}}\ne\infty$)
becomes significant, especially for high-order moments $q$.
This is because in  actual measurements
the NFMs at small bin size are 
determined  by the first few terms in 
the definition of the NFMs.
In most cases this leads  to a
significant underestimate of
the  measured NFMs with respect to their true values.

Cumulants are a more sensitive statistical tool (see \cite{rev}
and references therein).
However, their measurement is rather difficult and was rarely attempted. 
Besides, the cumulants are 
expected to be influenced by the 
statistical bias to even larger degree,
since they are constructed from the factorial moments 
of different orders $q$. 

\section{LOCAL PROPERTIES }

An important step towards an improvement of experimental measurements of
the local multiplicity distribution  was  made in \cite{bp1,bp2}, where
it was shown that any complex distribution can be represented as
$$
P_n(\delta )=P_0(\delta )\frac{\lambda^n}{n!}\, L_n, \qquad
L_n=\prod^{n}_{i=2}\eta_i^{n-i+1}(\delta ),
$$
where $\lambda =P_1(\delta)/P_0(\delta )$. 
The factor $L_n$ measures  a deviation of the distribution
from a Poisson, for  which  $L_n=1$. Non-poissonian fluctuations, therefore,
exhibit themselves as a deviation of $L_n$ from unity. The $L_n$   
are  constructed from the bunching parameters (BPs) 
\beq
\eta_q(\d )=\frac{q}{q-1}\frac{P_q(\d )P_{q-2}(\d )}
{P_{q-1}^2(\d )}, \qquad q>1. 
\label{l3}
\eeq
The expressions for the BPs and NFMs for 
some  popular distributions are
shown in Table~\ref{tab1}. The most
interesting observation is that while
the NFM  is an ``integral'' characteristic of the $P_n(\d )$
and the BP is a ``differential'' one, both tools
have values larger than unity  if the  distribution
is broader than a Poisson. 
Generally, however, one should not expect that all
BPs are larger than unity for a  broad distribution;
BPs probe  the distribution locally, i.e. they are simply
determined by the second-order derivative of  
the logarithm of $P_n(\d )$ with respect to $n$. 
Note that in the case of local distributions, 
the width of the distribution is mainly determined by 
$\eta_2(\d )$.
This observation is based on the
simple fact  that $P_n(\d )$  ceases to be bell-shaped at
sufficiently small $\d$.

BPs are more sensitive to the variation 
in the shape of $P_n(\d )$ 
with decreasing $\d$ than are NFMs \cite{che}. 
In the case of intermittent fluctuations,
one  should expect $\eta_2(\d )\propto\d^{-d_2}$.
For  multifractal local fluctuations, the $\eta_q(\d )$ are 
$\d$-dependent functions  for all $q\ge 3$,
while for   monofractal behaviour 
$\eta_q(\d ) =const$ for $q\ge3$ \cite{bp1}.

From an experimental point of view,
the BPs have the following important advantages \cite{bp2}:

\medskip
1) They are less severely affected by
the bias from finite statistics
than the NFMs, since the $q$th-order BP
resolves only the behaviour
of the multiplicity distribution near multiplicity $n=q-1$;

2)  For the calculation of the BP of order $q$,
one needs to know only the $q$-particle
resolution of the detector,
not any higher-order resolution.

\medskip
In this paper, we present an experimental
investigation of
local fluctuations in the final-state hadron system produced
in $Z^0$ decays at $\sqrt{s}=91.2$ GeV. 
The  final-state charged hadrons have been
recorded with the L3 detector during
the 1994 LEP running period.
The calculations are based
on approximately 1.0M  selected hadronic events.
We compare the data with 
the JETSET 7.4 PS \cite{eer9}, ARIADNE 4.08 
\cite{ard1} and HERWIG 5.9 \cite{h56}
models. The models have been tuned  to  the
L3 data  \cite{eer10}.

\section{ANALYSIS}
\label{sec:ex}

1) {\it Horizontal BPs}:

In order  to reduce
the statistical error on the observed local quantities
when analyzing experimental data,
we use the bin-averaged BPs \cite{bp1,bp2}:
\begin{equation}
\eta_q(M)=
\frac{q}{q-1} \frac{\bar N_q(M) \bar N_{q-2}(M)}
{\bar N_{q-1}^2(M)}, 
\label{l14}
\end{equation}
where $\bar N_{q}(M)=\frac{1}{M}\sum_{m=1}^{M}N_q(m,\delta )$,
$N_q(m,\delta )=$ being  the number of events  having $q$ particles
in  bin $m$ and $M=\Delta /\delta$  is the total number of bins
($\Delta$ represents the size of full phase-space volume). 
To be able to study  non-flat distributions, as  for rapidity, 
we have to carry out a transformation from the original
phase-space  variable to one 
in which the underlying
density is approximately uniform, as suggested by Bia\l as, 
Gadzinski and Ochs \cite{tr}. 

\medskip
2) {\it Generalized integral BPs:}

To study the distribution for spikes,
we will  consider the generalized integral BPs \cite{bp2} 
using  the  squared pairwise
four-momentum difference $Q^2_{12}=-(p_1-p_2)^2$.
In this variable, the definition of the BPs is given by
\beq
\chi_{q}(Q^2)=
\frac{q}{q-1} \frac{\Pi_q(Q^2) \Pi_{q-2}(Q^2)}
{\Pi^2_{q-1}(Q^2)},
\label{l15}
\eeq
where $\Pi_q(Q^2)$ represents the number
of events having  $q$  spikes of size 
$Q^2$ in the phase-space of variable $Q^2_{12}$ , 
irrespective of
how many particles are inside each spike.

\begin{figure}[htb]
\begin{center}~
\begin{picture}(100,100)
\put(0,10){
         {\epsfig{file=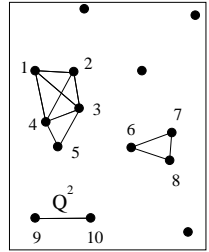, width=0.35\linewidth}}
         }
\end{picture}
\end{center}
\vspace{-1.5cm}

\caption{Example for the use of the Grassberger-Hentschel-Procaccia
counting topology in two dimensions.
Full points illustrate the position of particles in  an event.
Points whose interparticle distance is smaller than or equal to
the maximum $Q^2$, are connected by lines.
This configuration corresponds to
four spikes: one four-particle (1,2,3,4),
two three-particle (3,4,5), (6,7,8)  and
one two-particle  (9,10) spikes.}
\label{an1}
\vspace{-0.7cm}
\end{figure}

To define the spike size, we  shall use
the so-called Grassberger-Hentschel-Procaccia
counting topology
for which a many-particle hyper-tube is assigned
a size $Q^2$ corresponding to the
maximum of all pairwise distances (see Fig.~\ref{an1}).
For a Poissonian production,  
the BPs (\ref{l15})
are equal to unity for all $q$.

\subsection{In the rapidity variable}

\begin{figure}[htbp]
\begin{center}

\vspace{-0.1cm}
\mbox{\epsfig{file=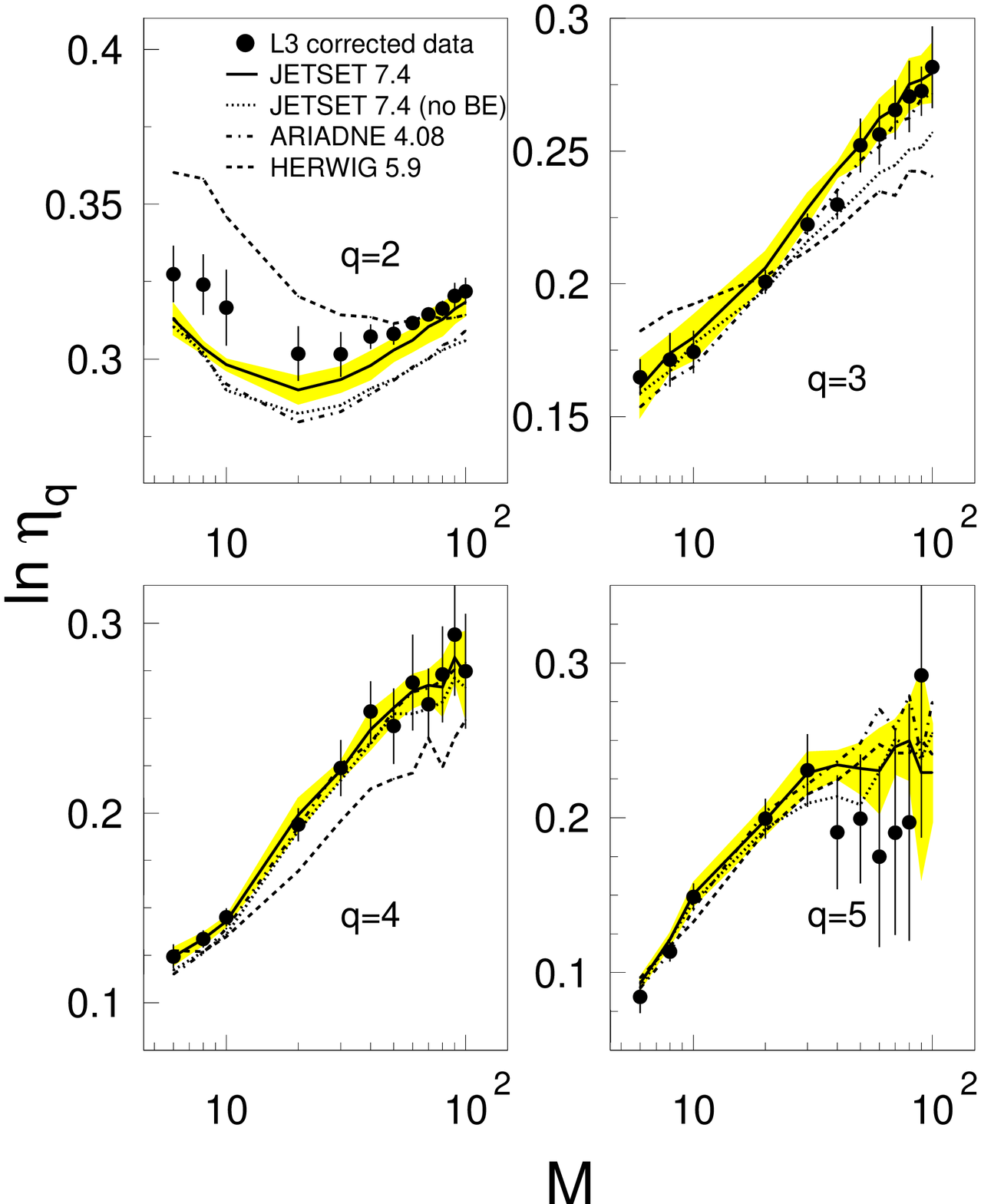, width=0.98\linewidth}}
\end{center}

\vspace{-1.0cm}
\caption{
BPs as a function of the number $M$ of bins in
rapidity defined with respect to the thrust axis.
The shaded areas represent the statistical and systematic errors
on the JETSET predictions.}
\label{bp_f}

\vspace{0.1cm}

\begin{center}
\mbox{\epsfig{file=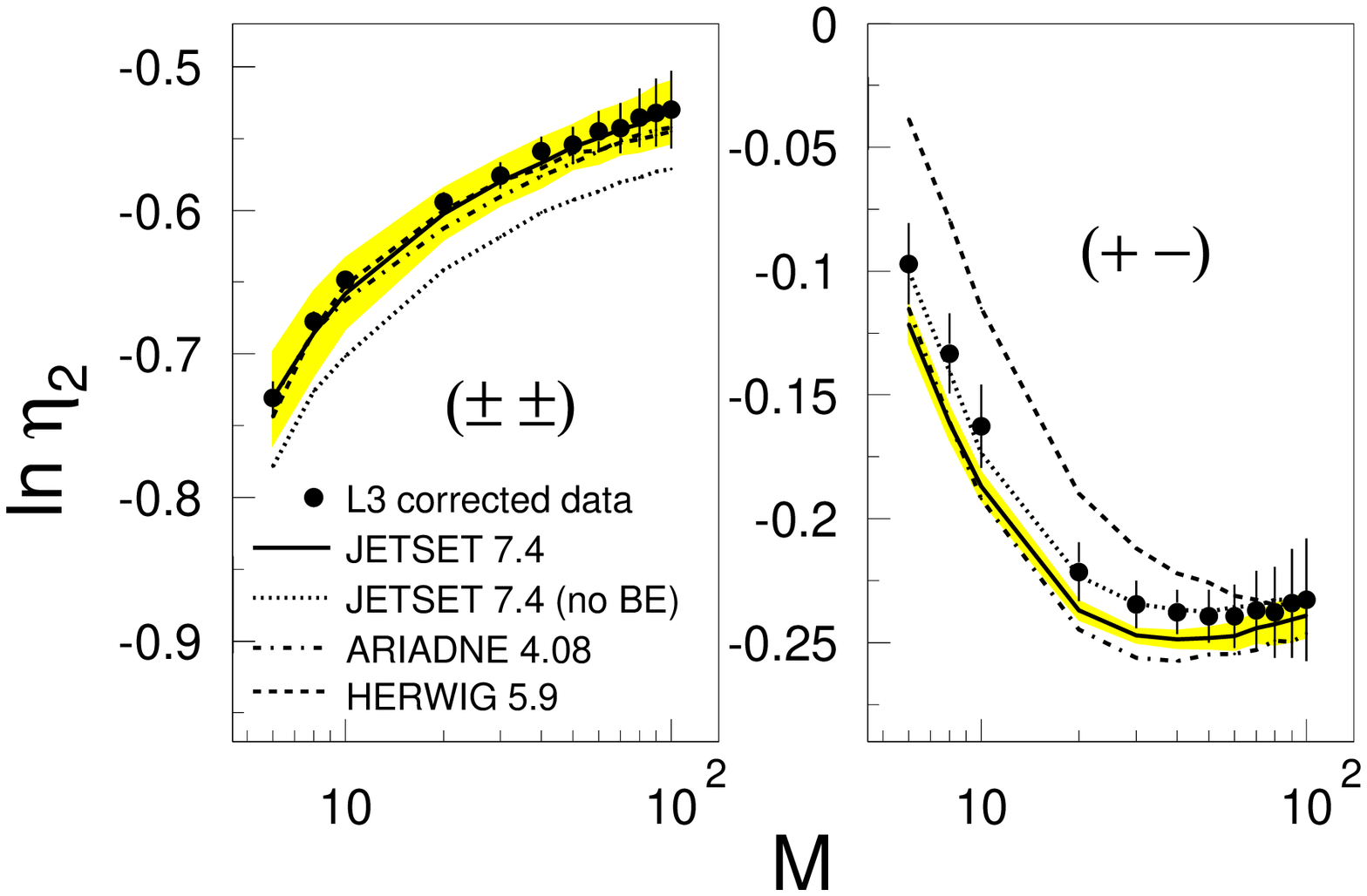, width=1.0\linewidth}}
\end{center}

\vspace{-1.0cm}
\caption{
The second-order BP  as a function of the number $M$
of bins in  rapidity defined with respect
to the thrust axis for
like-charged and unlike-charged particle combinations.
The shaded areas represent the errors
on the JETSET predictions.}
\label{bp_ff}
\end{figure}

In order to study  fluctuations inside jets,
in most investigations the 
fluctuations  have been measured in the rapidity
$y$ defined with respect to the thrust  
or sphericity axis \cite{cllr}.
The analysis for this variable is performed in the full
rapidity range $\mid Y \mid \leq 5$.
Fig.~\ref{bp_f}   shows the results for the 
BPs (\ref{l14}) for the  rapidity  variable  after
the Bia\l as-Gazdzicki-Ochs transformation.
The second-order BP  
decreases with increasing $M$ up to 
$M\simeq  20$, which is found to
correspond to the value of $M$ at which the maximum of
the multiplicity distribution $P_n(\delta)$
first occurs at $n=0$.
At large $M$, all  BPs show a
power-law  increase 
with increasing $M$, $\eta_q\sim M^{\alpha_q}$.
This indicates that the fluctuations  in $y$ defined
with respect to the thrust axis
are   multifractal scale invariant. 
Note that the conclusion that 
fluctuations have  a multifractal structure is  possible without the
necessity of  calculating  the intermittency indices $\phi_q$.
In contrast, to reveal multifractality with the help of the NFMs,
one first needs to carry out fits of the NFMs by a power law.

Both JETSET and ARIADNE agree well with the data for the higher orders.
However, ARIADNE underestimates $\eta_2$ for 
all bin sizes while JETSET is significantly
too low only for wide bins.
HERWIG predictions (dashed lines) significantly overestimate  the
second-order BP obtained from the data. Since 
the second-order BP is determined by  the dispersion
of the distribution,
this means that the HERWIG produces  too broad
local multiplicity distributions.
Such a result confirms that obtained by
the ALEPH Collaboration \cite{AL2}.

To study the disagreement   in
more detail, we split $\eta_2$ into two BPs:
\beq
\eta_2 = \eta_2^{(\pm\pm)} + \eta_2^{(+-)}.
\label{l3aa}
\eeq
Here, $\eta_2^{(\pm\pm)}$ is defined by (\ref{l14}) with
$N_2(m, \d)=N_2^{(\pm\pm)}(m, \d y)$, where $N_2^{(\pm\pm)}(m, \d y)$
is the number of events  having like-charged two-particle
combinations inside bin $m$ of size $\d y$.
Analogously, $\eta_2^{(+-)}$ is constructed from
the number of events $N_2^{(+-)}(m, \d y)$  having  unlike-charged
two-particle combinations. Note that due to
a combinatorial reason,
$\eta_2^{(\pm\pm)}<\eta_2^{(+-)}$.

Fig.~\ref{bp_ff}  shows that $\eta_2^{(\pm\pm)}$ and
$\eta_2^{(+-)}$ indeed behave completely differently.
While $\eta_2^{(\pm\pm)}$ has  the expected rise,
$\eta_2^{(+-)}$ shows a strong decrease at low $M$ and
an  onset of increase only at large $M$. The structure of
$\eta_2$ observed in Fig.~\ref{bp_f}  is
a combination of these two effects.

Let us remind that, in order to model the BE  interference in JETSET,    
the momenta of identical final-state
particles are shifted to reproduce the expected two-particle correlation
function. The main disadvantage of 
such an {\em ad hoc}  method is that it  
spoils  the overall energy-momentum conservation 
thus makes it necessary to modify also the momenta of non-identical
particles to compensate for this. This effect  
can be seen in Fig.~\ref{bp_ff}: 
JETSET without the BE correlations agrees with 
unlike-charged particles. 
However, JETSET with the BE effect disagrees with the data for 
unlike-charged
particle combinations, while it agrees with like-charged.

The strong anti-bunching tendency   seen for
unlike-charged particles at  $M <30$ can
be attributed to resonance decays and
to chain-like particle production  along the thrust axis,
as expected from the QCD-string model \cite{oder1}.
The latter effect leads to local charge conservation with
an alternating charge structure.
Evidence for this  effect was recently observed
by DELPHI \cite{oder2}.
As a result, there is  a  smaller
rapidity separation between
unlike-charged particles than between like-charged and
$\eta_2^{(+-)}$ is much larger than $\eta_2^{(\pm\pm)}$
at small $M$.
Having correlation lengths $\d y\sim 0.5-1.0$ in rapidity,
the resonance and the charge-ordering
effects, however,
become smaller with increasing $M$. 

\subsection{In the four-momentum difference}

The study of BPs described above can help us to understand 
a tendency of the particles to be grouped  into spikes inside small
phase-space intervals. Another question is
how the multiplicity of these spikes is  distributed from event to
event if the spike size goes to zero. 
To study this, we will use the BPs defined in (\ref{l15}).

\begin{figure}[htbp]
\begin{center}

\vspace{-0.2cm}
\mbox{\epsfig{file=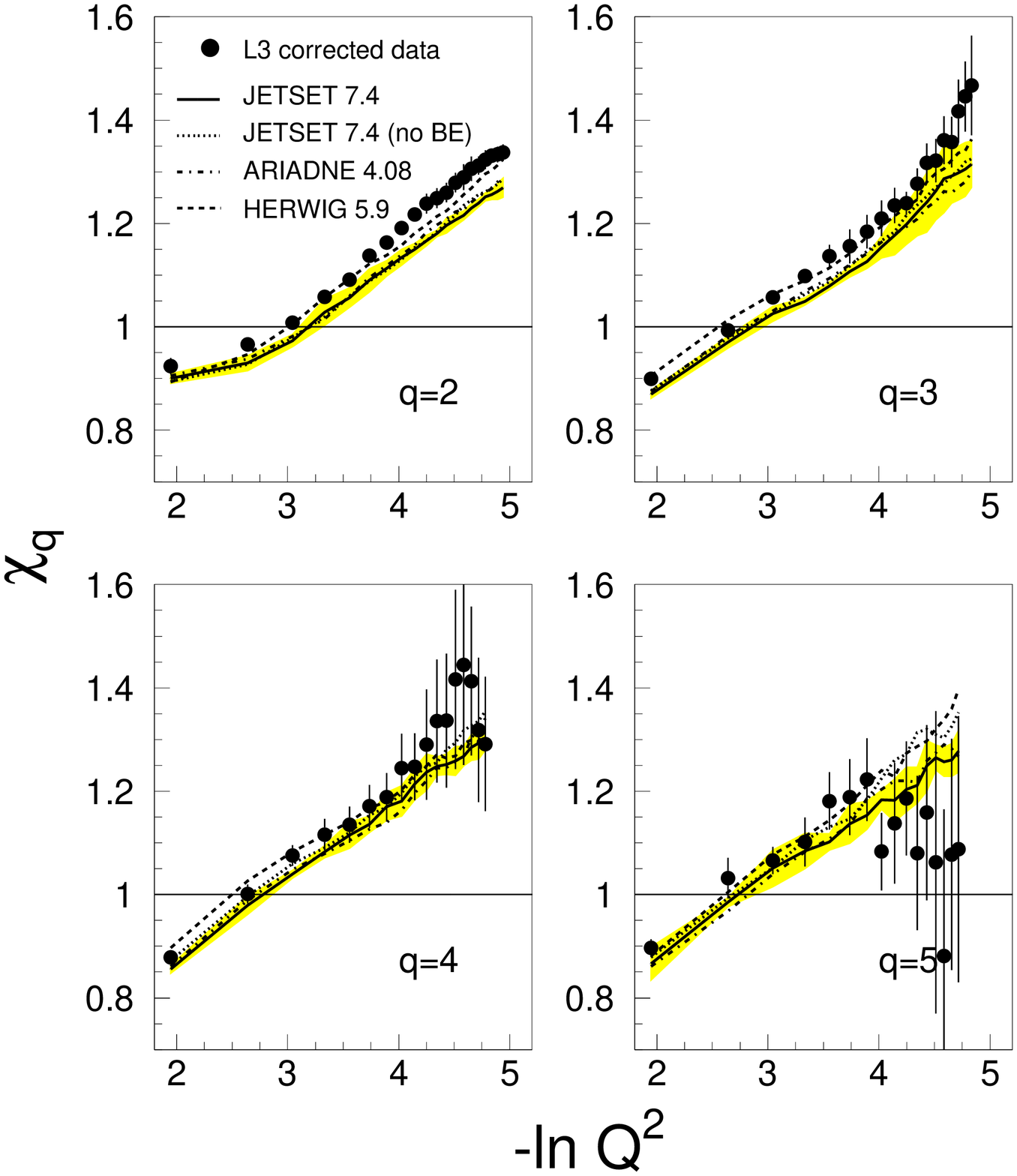, width=1.0\linewidth}}

\vspace{-1.0cm}
\caption{Generalized integral BPs
as a function of
the squared four-momentum difference $Q^2$
between two charged particles.
The shaded areas represent the errors
on the JETSET predictions.
}
\label{bpq_f}
\end{center}

\vspace{1.6cm}

\begin{center}
\mbox{\epsfig{file=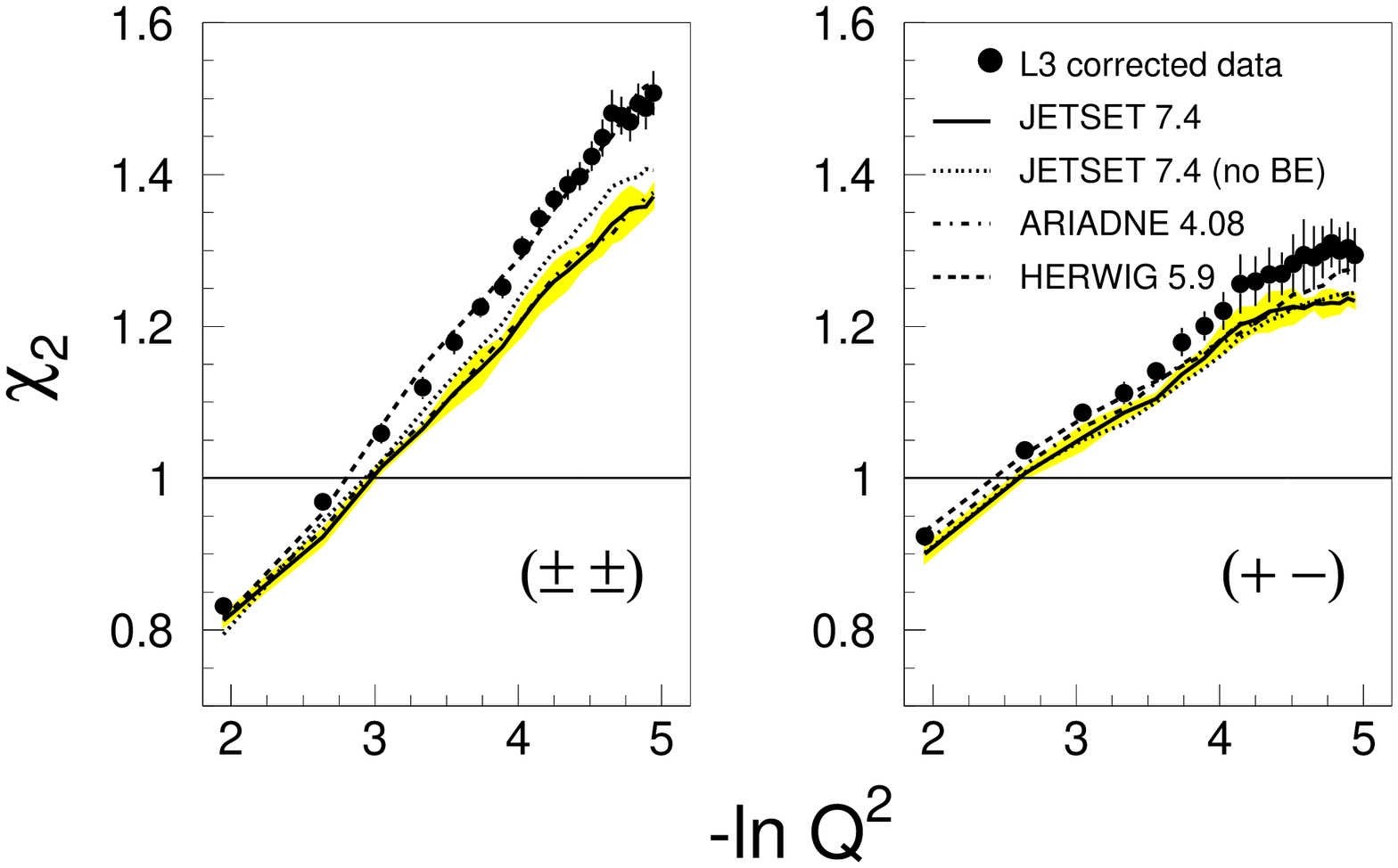, width=1.0\linewidth}}
\end{center}

\vspace{-1.0cm}

\caption{Generalized second-order BP
as a function of
the squared four-momentum difference $Q^2$
between two charged particles.
The shaded areas represent the errors
on the JETSET predictions.}
\label{bpq_ff}
\end{figure}

Fig.~\ref{bpq_f}  shows the behaviour of $\chi_q$
as a function of $-\ln Q^2$.
The  full lines represent the behaviour of the  BPs
in the Poissonian case.
In contrast, all BPs obtained from the  Monte Carlo models
rise with increasing $-\ln Q^2$ 
(decreasing $Q^2$).
This corresponds to a
strong bunching effect of all orders, as expected
for multifractal fluctuations.
The anti-bunching effect ($\chi_q<1$) for small $-\ln Q^2$ is
caused by the energy-momentum  conservation constraint \cite{bp2}.

To learn more about the mechanism of  multiparticle fluctuations
in  the variable $Q^2_{12}$, we present in Fig.~\ref{bpq_ff}
the behaviour of the second-order
BP as a function of $-\ln Q^2$ for
multiparticle  hyper-tubes (spikes) made of
like-charged and  those of  unlike-charged particles, separately.
For this kind of observables, HERWIG gives better agreement than
the LUND models. 

\section{DISCUSSION}
\label{disc}

Local multiplicity
fluctuations  were 
studied by means of  bunching parameters.
Since all high-order BPs show a power-like 
rise  with decreasing  size of phase-space interval, 
none of the conventional  multiplicity distributions
given  in Table~\ref{tab1} can describe the observed local 
fluctuations.

For  $\E$ interactions, one can be confident
that, at least on the parton level of this reaction,
perturbative QCD   can give a
hint for the understanding of the problem.
Analytical calculations based on the DLLA 
of perturbative QCD show that the multiplicity distribution
of partons in ever smaller opening angles is
inherently multifractal \cite{qcdd1}.
Qualitatively, this is consistent with our results on the  BPs 
for rapidity. 
Quantitatively, however, the QCD predictions
disagree with the $\E$ data and MC models \cite{qcd1}.

In this paper we show that the power-law behaviour of BPs
is mainly due to  like-charged particles.
JETSET gives the same power-law trend even without
the BE effect. This means that the  intermittency observed for
like-charged particles appears to be largely  a consequence of 
QCD parton showers and hadronization. 

The predictions of the ARIADNE 4.08 model are comparable with those
of the JETSET 7.4 PS model. This is essentially
due to the same implementation
of  hadronization,  which is based for both models
on string  fragmentation.

A noticeable disagreement is
found between the data and HERWIG model 
for rapidity variable.
The conversion of the partons into hadrons
in LUND  models is based on the Lund String Model \cite{oder1}.
However, the hadronization in HERWIG is 
modelled with a cluster mechanism \cite{h56}.
This  difference   can be a rather natural 
candidate to explain the observed disagreement.
A particular concern is the large discrepancy 
for $\eta_2$. The behaviour of $\eta_2$ 
for not very small intervals  is sensitive to  low-multiplicity events,
for which  hadronization details could play a significant role.

\section*{Acknowledgments}
This work is part of the research program of the ``Stichting voor
Fundamenteel Onderzoek der Materie (FOM)'', which
is financially supported by the ``Nederlandse Organisatie voor
Wetenschappelijk Onderzoek (NWO)''.
We acknowledge the effort of all engineers and
technicians who have participated
in the construction and 
maintenance of the LEP machine and the L3
detector.

\end{document}